\documentclass[aps,prl,twocolumn,groupedaddress,showpacs,amsmath]{revtex4}

\usepackage{graphicx}
\usepackage{amssymb}
\usepackage{comment}

\include{Letter.bbl}

\newcommand*{\br}{\mathbf{r}}

\newcommand*{\cF}{{\cal F}}

\newcommand*{\cH}{{\cal H}}

\newcommand*{\phop}{\phi^{\vphantom{\dagger}}}
\newcommand*{\phdop}{\phi^\dagger}
\newcommand*{\pop}{\pi^{\vphantom{\dagger}}}
\newcommand*{\pdop}{\pi^\dagger}
\newcommand*{\aop}{a^{\vphantom{\dagger}}}
\newcommand*{\adop}{a^\dagger}
\newcommand*{\bop}{b^{\vphantom{\dagger}}}
\newcommand*{\bdop}{b^\dagger}

\newcommand*{\zop}{z^{\vphantom{\dagger}}}
\newcommand*{\zdop}{z^\dagger}

\newcommand{\vect}[1]{\mathbf{#1}}

\begin{document}

\title{Quantum quenches in a spinor condensate}

\author{Austen Lamacraft}
\affiliation{Rudolf Peierls Centre for Theoretical Physics, 1 Keble Road, Oxford OX1 3NP, UK
and All Souls College, Oxford.}
\date{\today}
\email{a.lamacraft1@physics.ox.ac.uk}

\begin{abstract}

We discuss the ordering of a spin-1 condensate when quenched from its paramagnetic phase to its 
ferromagnetic phase by reducing magnetic field. We first elucidate the nature of the equilibrium 
quantum phase transition. Quenching rapidly through this transition reveals XY ordering either at a specific wavevector, or the `light-cone' correlations familiar from relativistic theories, depending on the endpoint of the quench. For a quench proceeding at a finite rate the ordering scale is governed by the Kibble-Zurek mechanism. The creation of vortices through growth of the magnetization fluctuations is also discussed. The long 
time dynamics again depends on the endpoint, conserving the order parameter in zero field, but not at 
finite field, with differing exponents for the coarsening of magnetic order. The results are discussed in the light of a recent experiment by Sadler \emph{et al.}

\end{abstract}


\maketitle

How does a many-particle system undergo condensation into an ordered state? This question is central 
to a number of disparate areas of physics, from condensed matter to cosmology~\cite
{bray1994,volovik2003}. Often we are interested in the processes determining the 
formation of ordered domains and topological defects. The usual approach is to study the coupled 
dynamics of the collective (or hydrodynamic) degrees of freedom, including the order parameter and any 
conserved quantities. Thus the dynamics is highly constrained by the presence or absence of 
conservation laws, with dramatic differences in the resulting time evolution of correlations. Such considerations
will be important when we quench into an ordered state at zero temperature through a \emph{quantum} 
phase transition~\cite{cherng2006,calabrese2006}. 

Such a possibility was explored in a recent experiment that studied ferromagnetic ordering in a Bose-Einstein condensate of $^{87}$Rb atoms following a sudden reduction in magnetic field~\cite{sadler2006}. Cold atomic gases represent an exciting new prospect for the investigation of such quantum quenches. 
As we will show, they represent a far closer analog of relativistic theories than the condensed matter 
systems suggested for `laboratory cosmology' by Zurek and reviewed in Ref.~\cite{zurek1996}


Earlier, mostly numerical work~\cite{pu1999,robins2001,saito2005,zhang2005,mur-petit2006} 
 has focused on treating the creation of spin domains in condensates as a property of a classical dynamical 
system. 
Our goal in this Letter, on the other 
hand, is to first explain the \emph{character} of the equilibrium quantum phase transition, and then to 
discuss the associated dynamics as a problem of \emph{phase ordering}. In particular, this will lead us 
to carefully distinguish different quenches in terms of the conservation laws obeyed, and the resulting 
dynamics of topological defects (vortices in the magnetization). These are features not present in the model systems discussed in Refs.~\cite{cherng2006,calabrese2006}, for instance, and their treatment requires the introduction of some novel theoretical ideas that should be of broad applicability. 


%
\begin{figure} 
\centering  \includegraphics[width=0.43\textwidth]{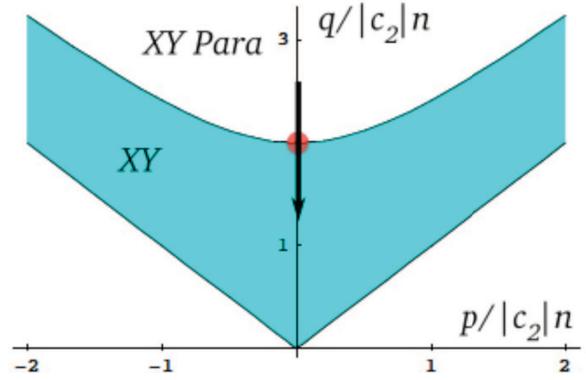}
\caption{Zero temperature phase diagram of a spin-1 condensate in terms of the linear and quadratic Zeeman energies. The shaded area corresponds to the region of XY ordering. The experiment of Ref.~\cite{sadler2006}  involved a quench through the special point at $p=0$ (red dot) from the paramagnetic region.\label{fig:pd}}
\end{figure}

The existence of an ordering transition in the ferromagnetic spin-1 Bose gas is readily understood on the 
basis of a variational Gross-Pitaevskii calculation~\cite{stenger1998}. The second-quantized 
Hamiltonian is~\cite{ho1998,ohmi1998}
\begin{eqnarray}\label{Ham}
H&=&\int d\br\, \cH_{\mathrm{0}}+\cH_{\mathrm{Int}}\\
\cH_{0}&=&\phdop_m \left[-\frac{1}{2}\nabla^2+E^\mathrm{Z}_m\right]\phop_m
\nonumber\\
\cH_{\mathrm{Int}}&=&\frac{1}{2}\phdop_m\phdop_{m'}\left[c_0\delta_{mn}\delta_{m'n'} 
+c_2\vect{F}_{mn}\cdot\vect{F}_{m'n'}\right]\phop_{n'}\phop_{n}\nonumber
\end{eqnarray}
The index $m=-1,0,1$ gives the z-component of total spin of the corresponding state, and we have set $
\hbar$ and the atomic mass to unity. $F^i_{mn}$ are the spin-1 matrices for $i=x,y,z$. 
$E^\mathrm{Z}_m$ is the Zeeman energy of the $m$-component, defined below. For a ferromagnetic system the spin interaction parameter $c_2$ is negative.
We implement the Gross-Pitaevskii approximation by 
treating the $\phop_m$ as c-numbers $\varphi_m$ and writing $\varphi_m=\sqrt{n}\hat\chi_m
$ in terms of a normalized spinor $\hat\chi_m$. The energy per particle is  
%
\pagebreak
\begin{widetext}
\begin{eqnarray}\label{egp} 
\frac{c_0n}{2}+\frac{c_2n}{2}\left[\left(|\chi_1|^2-|\chi_{-1}|^2\right)^2+2\left(|
\chi_0\chi_1|^2+|\chi_0\chi_{-1}|^2+\chi_0^{*2}\chi_1\chi_{-1}+\chi_0^2
\chi^*_1\chi_{-1}^*\right)\right]
+q\left[|\chi_1|^2+|\chi_{-1}|^2\right]-p\left[|\chi_1|^2-|\chi_{-1}|^2\right].\nonumber\\
\end{eqnarray}
\end{widetext}
The last two terms originate from the linear and quadratic Zeeman energies, the general case for spin-1:
$E^\mathrm{Z}_m\equiv-\tilde p m+qm^2$.
%
In fact the coefficient $p$ in the energy is the sum of the linear Zeeman term and a Lagrange 
multiplier enforcing conservation of $F_z$. We should minimize Eq.~(\ref{egp}) and then use the 
true value of the magnetization $\langle F_z\rangle$ to fix $p$. Thus with zero total magnetization we 
will have $p=0$. In this case it is straightforward to see that for $q>q_0\equiv 2|c_2|n$ the spinor state $\hat\chi^
{\dagger}=(0,1,0)$ minimizes the energy, while for $q<q_0$ the $\pm 1$ states become populated, 
leading to a transverse magnetization density $f_{\perp}(\br)\equiv f^x(\br)+if^y(\br)$, where $f^i(\br)\equiv\phdop_m(\br) F^i_{mn}\phop_n(\br)$. In the general case $p\neq 0$~\cite{murata2006}
\begin{equation}\label{Ftrans}
|\langle f_\perp\rangle|=n\frac{\sqrt{q^2-p^2}\sqrt{\left(p^2+q_0q\right)^2-q^4}}{q_0q^2},
\end{equation}
showing that $\langle f_{\perp}\rangle$ is nonzero between the lines $|p|=q$ and $|p|=\sqrt{q(q-q_0)
}$ (see Fig.~\ref{fig:pd}). Thus the mean-field calculation predicts a symmetry-breaking quantum phase transition.
For $p\neq 0$, there is also a perpendicular magnetization in this region
\[\langle f_z\rangle=n\frac{p\left(p^2+q_0q-q^2\right)}{q_0q^2}.\]
%
and the ordered phase is a canted XY ferromagnet~\cite{mukerjee2006}, while the $f_\perp=0$ phase is an XY paramagnet (finite transverse susceptibility).

We now ask in more detail what kind of quantum phase transition we are dealing with. This is more than a formal question, as the dynamics of the order parameter at the transition will be crucial in determining the behavior following a quench. We start by considering the Bogoliubov theory of the paramagnetic phase. Shifting the fields $\phop_m$ in Eq.~(\ref{Ham}) by $\varphi^{\dagger}=(0,\sqrt{n},0)$ we find that in the quadratic part of the Hamiltonian the $\phop_{\pm 1}$ states decouple from the $\phop_0$ state to give 
\begin{eqnarray}\label{xyBog}
H_B\equiv\sum_{k,m=\pm 1} \phdop_{k,m}\left[\varepsilon_k-pm+q+c_2 n\right]\phop_{k,m}\nonumber\\
+c_2n\sum_{k} \phdop_{k,1}\phdop_{-k,-1}+\phop_{k,1}\phop_{-k,-1}.
\end{eqnarray}
($\varepsilon_k\equiv k^2/2$) This is readily diagonalized by Bogoliubov transformation to yield (except for a constant)
%
%
%
\begin{equation}\label{diag_p}
H_B=\sum_k E_{s,+}(k)\adop_k\aop_k+E_{s,-}(k)\bdop_k\bop_k,
\end{equation}
where $E_{s,\pm}(k)=E_s(k)\mp p$, with the spin wave dispersion $E^2_s(k)=\left(\varepsilon_k+q\right)\left(\varepsilon_k+q+2c_2 n\right)$. One of these dispersions passes through zero when $|p|=p_c\equiv \sqrt{q(q-q_0)}$, the same instability of the paramagnetic phase that we found before.
%
%
%
%
%
%
%
%
%
%
%
Except at $p=0$ the transition to the ordered state proceeds by filling of either the `particle' or `hole' band in Eq.~(\ref{diag_p}), and involves a change in $F_z$. Since this is conserved, 
it is impossible to cross this transition without contact to a reservoir of magnetization (unless $T\neq 0$). When $p=0$, on the other hand, the transition occurs through closing of the bandgap and the longitudinal magnetization remains zero.

To make a connection to the general theory of quantum phase transitions, we rewrite the Hamiltonian Eq.~(\ref{diag_p}) using the (complex) canonical 
coordinates and conjugate momenta
\begin{equation*}
 \zop_k\equiv \frac{1}{\sqrt{E_s(k)}}\left(\aop_k+\bdop_{-k}\right),\,\pop_k\equiv i\sqrt{E_s(k)}\left(\adop_k-\bop_{-k}\right)
\end{equation*}
%
In this way we get (dropping the momentum sum)
\begin{equation}\label{Hp}
H_B= \frac{1}{2}\left( \pdop+ip z\right)\left( \pop-ipz^\dagger\right)+\frac{1}{2}\left[E_s^2-p^2\right] z^{\dagger} z.
\end{equation}
Eq.~(\ref{Hp}) is recognized as the Hamiltonian of a two-dimensional particle in a uniform perpendicular magnetic field and 
harmonic oscillator potential.

How do we interpret the field $z$? The Fourier modes of the transverse magnetization density $f_{\perp}(\br)$ may be written in terms of $z_k$ as
\begin{eqnarray*}
f_{\perp\,k}&=&\sqrt{2}\sum_{l} \phdop_{l+k,1}\phop_{l,0}+\phdop_{l+k,0}\phop_{l,-1} \nonumber\\
&=&\sqrt{2n(\varepsilon_k+q)}z^\dagger_k+\cdots
\end{eqnarray*}
where the dotted lines denote terms higher order in the quasiparticle operators. At low $k$ the two are simply proportional, as one would hope. 

Below the transition, the higher order terms dropped from the Hamiltonian Eq.~(\ref{xyBog}) are required to saturate the growth of $f_{\perp}(\br)$. Close enough to the transition a quartic term is sufficient. Obtaining this term within the Bogoliubov theory is slightly subtle as it involves the partial cancellation of the `direct' quartic interaction neglected in Eq.~(\ref{xyBog}) against the interaction induced by phonons~\cite{lamacraft2006}. The most relevant term may be found more simply, aside from a small renormalization, from the observation that it is responsible for the square root singularity in Eq.~(\ref{Ftrans}) in the mean-field approximation. We write the final result as an effective action valid near the transition, and for length scales $\gg q^{-1/2}$~\footnote{We ignore quartic terms that mix the field $z$ and the momentum $\pi$ as irrelevant.}.
%
\begin{eqnarray}\label{Seff}
S_{\mathrm{eff}}=\int d\br\,dt\, \frac{1}{2}\left[\dot z^{\dagger}\dot z-c_s^2\nabla z^{\dagger}\nabla z-\left(p_c^2-p^2\right)z^{\dagger}z\right]\nonumber\\
-ipz^\dagger\dot z-\frac{|\tilde c_2|q^2}{2}|z|^4
\end{eqnarray}
%
 $\tilde c_2\equiv(1-\frac{|c_2|}{c_0})c_2$. To obtain the quadratic part of Eq.~(\ref{Seff}) from Eq.~(\ref{Hp}) we have approximated the spectrum as
\begin{equation}\label{rel_spec}
E^2_{s}(k)\sim c_s^2k^2+p_c^2\qquad c_s^2\equiv q-q_0/2
\end{equation}
The precise form of the anharmonic terms in Eq.~(\ref{Seff}) will not be important in the following development.

$S_{\mathrm{eff}}$ is identical to the effective theory describing the superfluid-insulator transition in the Bose-Hubbard model. As in that problem, the point $p=0$ is identified as a special point where the transition, instead of being of the bose condensation type, lies in the universality class of the $(d+1)$-dimensional XY model~\cite{fisher1989}. It is relevant to ask whether the deviations from mean field critical behavior implied by this identification will be seen in experiment. For a two-dimensional condensate a standard calculation gives the Ginzburg criterion for the breakdown of mean field behavior~\cite{lamacraft2006}
$\frac{|\tilde c_2|}{\pi L_{\perp}}\left(\frac{q_0}{q-q_0}\right)^{1/2}\gtrsim1$
where $L_{\perp}$ is the transverse dimension. Since for the system in Ref.~\cite{sadler2006} the prefactor is of order $10^{-4}$, the mean field theory is an excellent approximation.

With this mind, we now proceed to describe the evolution of a system that is quenched suddenly through the transition at $p=0$ from an initial value $q_i>q_0$ for $t<0$ to $q_f<q_0$ at $t>0$,
as in Ref.~\cite{sadler2006}. There is a band of unstable modes with $E^2_s(k,q_f)<0$, the occupancy of which begins to grow exponentially, as they are populated with pairs of atoms scattering from the $m=0$ state. The quadratic Hamiltonian Eq.~(\ref{diag_p}) describes this process adequately until the populations are such that the anharmonic interactions between the modes -- such as the last term of Eq.~\ref{Seff} -- become important. Writing the Hamiltonian for $t>0$ in terms of the fields $\zop_k$ defined at $q=q_i$, and with $\omega_k^2\equiv-E^2_s(k,q_f)$, we find the solution of the Heisenberg equations of motion
\begin{eqnarray*}
\zop_k(t)&=&\zop_k(0)\cosh\omega_k t+\pop_k(0)\omega_k^{-1}\frac{\varepsilon_k+q_f}{\varepsilon_k+q_i}\sinh\omega_k t\nonumber\\
\end{eqnarray*}
The calculation of the correlation function of the transverse magnetization is then staightforward
\begin{eqnarray*}\label{mag_corr}
\langle f^{\vphantom{\dagger}}_{\perp k}(t)f^{\dagger}_{\perp -k}(t)\rangle=2n\left(\varepsilon_k+q_i\right)\left[\cosh^2\omega_k t\langle \zdop_k(0)\zop_k(0)\rangle\right.\nonumber\\
\left.+\omega_k^{-2}\left(\frac{\varepsilon_k+q_f}{\varepsilon_k+q_i}\right)^2\sinh^2\omega_k t\langle\pdop_k(0)\pop_k(0)\rangle\right],
\end{eqnarray*}
%
revealing the exponential growth of the magnetization. The initial fluctuations of the oscillator modes are~\footnote{Finite temperature can be included easily here through factors $\coth\left(E_s(k)/2k_BT\right)$.}
%
\[\langle \zdop_k(0)\zop_k(0)\rangle=(E_s(k,q_i))^{-1},\,\langle\pdop_k(0)\pop_k(0)\rangle=E_s(k,q_i).\]
%
These general formulae are valid for any instantaneous quench~\footnote{Note that the linear Zeeman term leads to a trivial Larmor precession of the magnetization that is not included.}. In the following we will make the simplification of taking $q_i\gg q_0$ (shot noise limit) , which gives
%
\begin{eqnarray*}
\langle f^{\vphantom{\dagger}}_{\perp k}(t)f^{\dagger}_{\perp -k}(t)\rangle=2n\left[\cosh^2\omega_k t
+\left(\frac{\varepsilon_k+q_f}{\omega_k}\right)^2\sinh^2\omega_k t\right].
\end{eqnarray*}
Now we wish to focus on two particular values of $q_f$ to illustrate the different possible classes of behavior. If $q_f=0$ the spectrum of unstable modes is $\omega_k^2=\varepsilon_k\left(q_0-\varepsilon_k\right)$, which has a maximum at $k=\sqrt{q_0}$.  The correlation function is therefore dominated by the fluctuations on this scale that grow at a rate $q_0$. Taking into account only the unstable modes, we find for the asymptotic behavior of the real space correlations
\begin{eqnarray*}
\langle f^{\vphantom{\dagger}}_{\perp}(\br,t)f_{\perp}^{\dagger}(\br',t)\rangle\to\frac{n}{2L_{\perp}}\sqrt{\frac{q_0}{2\pi t}}J_0(\sqrt{q_0}|\br-\br'|)e^{q_0t}
\end{eqnarray*}
for $q_0t\gg 1,r/t\ll c_s.$ The Bessel function is an angular average of plane waves of wavevector $q_{0}^{1/2}$. The result is a growing random spin texture of typical scale $q_{0}^{-1/2}$, as observed in Ref.~\cite{sadler2006}. Note that the vanishing of the mode growth rate at zero wavevector is a consequence of the conservation of all three spin components in zero field.

Very different behavior results if $q_f$ is only just below $q_0$, so that $p_c/c_s^2<<1$. In this case the spectrum of unstable modes reflects the relativistic form of Eq.~(\ref{rel_spec})
\[\omega_k^2=c_s^2\left(k_c^2-k^2\right),\]
with $k_c^2\equiv-p_c^2/c_s^2$ the `Compton wavevector'. In this case we get the asymptotic behavior
\begin{equation}\label{rel_asym}
\langle f^{\vphantom{\dagger}}_{\perp}(\br,t)f_{\perp}^{\dagger}(\br',t)\rangle\to \frac{nq_0}{4\pi L_{\perp}}\frac{1}{c_sk_c t}e^{k_c(4c^2_st^2-|\br-\br'|^2)^{1/2}},
\end{equation}
valid when the exponent is large. The correlation function Eq.~(\ref{rel_asym}) displays a striking growth of correlations along a `light cone' originating at a point halfway between $\br$ and $\br'$ and propagating at the spin wave velocity $c_s$. This is a familiar feature of spinodal instabilities in relativistic theories~\cite{weinberg1987}. The crossover between these two types of behavior occurs at the value $q_f=q_0/2$, where the maximum in the spectrum of unstable modes $k_\mathrm{max}\equiv\sqrt{q_0-2q}$ goes to zero. 

Next we discuss what happens if the quench is not instantaneous, but rather crosses the transition in some finite time. For concreteness we take $q(t)=q_0\left(1-t/\tau_Q\right)$, where $\tau_Q$ measures the duration of the quench, and the transition is crossed at $t=0$. The result may be obtained exactly in terms of Airy functions, but the following integral representation is more useful
\begin{eqnarray*}
\langle f^{\vphantom{\dagger}}_{\perp}(\br,t)f_{\perp}^{\dagger}(\br',t)\rangle=\frac{nq_0}{2c_s^2t_{\mathrm{KZ}}L_{\perp}}\cF(t/t_{\mathrm{KZ}},|\br-\br'|/(c_st_{\mathrm{KZ}}))\nonumber\\
\cF(x,y)\equiv\frac{1}{\pi^{3/2}}\int_0^{\infty} du\, u^{-3/2} \exp\left[ux-\frac{y^2}{4u}-\frac{u^3}{12}\right],
\end{eqnarray*}
%
where we have introduced $t_{\mathrm{KZ}}\equiv\left(\tau_Q/q_0^2\right)^{1/3}$. This expression can then be evaluated in the saddle-point approximation. At $|\br-\br'|=0$ we get an exponential factor $\exp\left(\frac{4}{3}\left(t/t_{\mathrm{KZ}}\right)^{3/2}\right)$. At this point we have to invoke for the first time the effect of the anharmonic interactions between modes. Very crudely, their effect is to cut-off the exponential growth of the magnetization. We shall not try to discuss this process in detail, but the key point is that it occurs at a time $\propto t_{\mathrm{KZ}}$, where the constant of proportionality may contain $t_{\mathrm{KZ}}$, but only logarithmically. Thus we readily see that the associated scale is
%
$\xi(t_{\mathrm{KZ}})\propto c_st_{\mathrm{KZ}}=c_s\left(\tau_Q/q_0^2\right)^{1/3}$.
%
This result is consistent with the general arguments of Kibble and Zurek, implying a domain size scaling as $(\tau_Q)^{\frac{\nu}{z\nu+1}}$, with mean field values $z=1$ and $\nu=1/2$ for the dynamic and correlation length exponents~\cite{zurek1996}.



The growth of the transverse magnetization is associated with the appearance of vortices. As the population of the unstable modes becomes large, the field $f^{\vphantom{\dagger}}_{\perp}(\br)$ can be treated as an effectively classical Gaussian stochastic variable, with variance given by the correlation functions calculated above~\cite{guth1985}. Then the density of vortices can be estimated using the Halperin-Liu-Mazenko formula to calculate the density of zeroes of this classical field $n_V(t)=-\frac{1}{2\pi}\partial_r^2 g''(r,t)|_{r=0}$~\cite{halperin1981,liu1992},
%
%
where $g(|\br-\br'|,t)\equiv \langle f^{\vphantom{\dagger}}_{\perp}(\br,t)f_{\perp}^{\dagger}(\br',t)\rangle/\langle f^{\vphantom{\dagger}}_{\perp}(0,t)f_{\perp}^{\dagger}(0,t)\rangle$ is the normalized correlation function. For quenches to $q_f<|c_2|n$ it is immediately clear that the density is determined by $k_\mathrm{max}$, as the spectrum of fluctuations is essentially monochromatic at late times, and $n_V(t)\to k_{\mathrm{max}}^2/4\pi$.
%
In this case the vortices have a core size of the same order as the scale of the magnetic order. For a quench to just below the transition, on the other hand, the asymptote Eq.~(\ref{rel_asym}) gives
%
$n_V(t)\to \frac{1}{4\pi} \frac{k_c}{c_s t}$.
%
This behavior continues until the growth is saturated by the anharmonic terms, which happens when $|f_{\perp}|^2\sim 2n\left(q_0-q\right)/\tilde c_2$. Finally, in the case of the finite time quench, we have $n_V\propto \xi^{-2}(t_{\mathrm{KZ}})$.

In closing, we discuss the long-time behavior of the system, once the transverse magnetization is comparable to its equilibrium value. This regime is characterized by the growth of the characteristic ordering scale and the annihilation of topological defects, usually called \emph{coarsening}. We distinguish two universality classes depending on whether or not the order parameter is conserved~\cite{bray1994}. In the first case the domain size increases as $\xi(t)\propto t^{1/2}$~\footnote{For a complex order parameter in $d=2$ the presence of vortices implies $\xi(t)\propto(t/\ln t)^{1/2}$.}, while in the second a $t^{1/4}$ law is obeyed. In our system these two cases correspond to a final value $q_f\neq 0$ or $q_f=0$ respectively. Note that for the shallow quench we found the $t^{1/2}$ behavior already at the linear level.

A further complication is that coarsening is usually studied using models of dissipative dynamics, where energy is not conserved. On the other hand,  coarse-graining of a purely Hamiltonian system can give rise to such dynamics, at the expense of introducing a conserved energy density to which the order parameter is coupled~\cite{hohenberg1977}. For the case of a real scalar non-conserved order parameter, Hamiltonian coarsening was examined Ref.~\cite{kockelkoren2002}, with the conclusion that the $t^{1/2}$ law was preserved (this case corresponds to Model C in the classification of Ref.~\cite{hohenberg1977}). On the other hand, Ref.~\cite{damle1996} studied coarsening in the Gross-Pitaevskii equation, where energy and additionally particle number are conserved (Model F), and found results consistent with $\xi(t)\propto t$. The dynamics described by Eq.~(\ref{Seff}) corresponds to Model F, except at the particle-hole symmetric point $p=0$ that has been our main concern, a special case called Model E. 

I would like to thank John Chalker, Joel Moore,  Sabrina Leslie, and Julien Kockelkoren for useful discussions. After this work was finished, the preprint Ref.~\cite{saito2006b} appeared, where the same experiment is discussed.


\begin{thebibliography}{26}
\expandafter\ifx\csname natexlab\endcsname\relax\def\natexlab#1{#1}\fi
\expandafter\ifx\csname bibnamefont\endcsname\relax
  \def\bibnamefont#1{#1}\fi
\expandafter\ifx\csname bibfnamefont\endcsname\relax
  \def\bibfnamefont#1{#1}\fi
\expandafter\ifx\csname citenamefont\endcsname\relax
  \def\citenamefont#1{#1}\fi
\expandafter\ifx\csname url\endcsname\relax
  \def\url#1{\texttt{#1}}\fi
\expandafter\ifx\csname urlprefix\endcsname\relax\def\urlprefix{URL }\fi
\providecommand{\bibinfo}[2]{#2}
\providecommand{\eprint}[2][]{\url{#2}}

\bibitem[{\citenamefont{Bray}(1994)}]{bray1994}
\bibinfo{author}{\bibfnamefont{A.~J.} \bibnamefont{Bray}},
  \bibinfo{journal}{Adv. Phys.} \textbf{\bibinfo{volume}{43}},
  \bibinfo{pages}{357} (\bibinfo{year}{1994}).

\bibitem[{\citenamefont{Volovik}(2003)}]{volovik2003}
\bibinfo{author}{\bibfnamefont{G.~E.} \bibnamefont{Volovik}},
  \emph{\bibinfo{title}{The Universe in a Helium Droplet}}
  (\bibinfo{publisher}{Clarendon Press}, \bibinfo{year}{2003}).

\bibitem[{\citenamefont{Cherng and Levitov}(2006)}]{cherng2006}
\bibinfo{author}{\bibfnamefont{R.~W.} \bibnamefont{Cherng}} \bibnamefont{and}
  \bibinfo{author}{\bibfnamefont{L.~S.} \bibnamefont{Levitov}},
  \bibinfo{journal}{Phys. Rev. A} \textbf{\bibinfo{volume}{73}},
  \bibinfo{pages}{043614} (\bibinfo{year}{2006}).

\bibitem[{\citenamefont{Calabrese and Cardy}(2006)}]{calabrese2006}
\bibinfo{author}{\bibfnamefont{P.}~\bibnamefont{Calabrese}} \bibnamefont{and}
  \bibinfo{author}{\bibfnamefont{J.}~\bibnamefont{Cardy}},
  \bibinfo{journal}{Phys. Rev. Lett.} \textbf{\bibinfo{volume}{96}},
  \bibinfo{pages}{136801} (\bibinfo{year}{2006}).

\bibitem[{\citenamefont{Sadler et~al.}(2006)\citenamefont{Sadler, Higbie,
  Leslie, Vengalattore, and Stamper-Kurn}}]{sadler2006}
\bibinfo{author}{\bibfnamefont{L.~E.} \bibnamefont{Sadler \emph{et al.}}},
 \bibinfo{journal}{Nature}
  \textbf{\bibinfo{volume}{443}}, \bibinfo{pages}{312} (\bibinfo{year}{2006}).

\bibitem[{\citenamefont{Zurek}(1996)}]{zurek1996}
\bibinfo{author}{\bibfnamefont{W.~H.} \bibnamefont{Zurek}},
  \bibinfo{journal}{Physics Reports} \textbf{\bibinfo{volume}{276}},
  \bibinfo{pages}{177} (\bibinfo{year}{1996}).

\bibitem[{\citenamefont{Pu et~al.}(1999)\citenamefont{Pu, Law, Raghavan,
  Eberly, and Bigelow}}]{pu1999}
\bibinfo{author}{\bibfnamefont{H.}~\bibnamefont{Pu \emph{et al.}}},
 \bibinfo{journal}{Phys. Rev. A}
  \textbf{\bibinfo{volume}{60}}, \bibinfo{pages}{1463} (\bibinfo{year}{1999}).

\bibitem[{\citenamefont{Robins et~al.}(2001)\citenamefont{Robins, Zhang,
  Ostrovskaya, and Kivshar}}]{robins2001}
\bibinfo{author}{\bibfnamefont{N.~P.} \bibnamefont{Robins \emph{et al.}}},
 \bibinfo{journal}{Phys. Rev. A}
  \textbf{\bibinfo{volume}{64}}, \bibinfo{pages}{021601(R)}
  (\bibinfo{year}{2001}).

\bibitem[{\citenamefont{Saito and Ueda}(2005)}]{saito2005}
\bibinfo{author}{\bibfnamefont{H.}~\bibnamefont{Saito}} \bibnamefont{and}
  \bibinfo{author}{\bibfnamefont{M.}~\bibnamefont{Ueda}},
  \bibinfo{journal}{Phys. Rev. A} \textbf{\bibinfo{volume}{72}},
  \bibinfo{eid}{023610} (\bibinfo{year}{2005}).

\bibitem[{\citenamefont{Zhang et~al.}(2005)\citenamefont{Zhang, Zhou, Chang,
  Chapman, and You}}]{zhang2005}
\bibinfo{author}{\bibfnamefont{W.}~\bibnamefont{Zhang \emph{et al.}}},
  \bibinfo{journal}{Phys. Rev. Lett} \textbf{\bibinfo{volume}{95}},
  \bibinfo{eid}{180403} (\bibinfo{year}{2005}).

\bibitem[{\citenamefont{Mur-Petit et~al.}(2006)\citenamefont{Mur-Petit,
  Guilleumas, Polls, Sanpera, Lewenstein, Bongs, and
  Sengstock}}]{mur-petit2006}
\bibinfo{author}{\bibfnamefont{J.}~\bibnamefont{Mur-Petit \emph{et al.}}},
  \bibinfo{journal}{Phys. Rev. A} \textbf{\bibinfo{volume}{73}},
  \bibinfo{eid}{013629} (\bibinfo{year}{2006}).

\bibitem[{\citenamefont{Stenger et~al.}(1994)\citenamefont{Stenger, Inouye,
  Stamper-Kurn, Miesner, Chikkatur, and Ketterle}}]{stenger1998}
\bibinfo{author}{\bibfnamefont{J.}~\bibnamefont{Stenger \emph{et al.}}},
  \bibinfo{journal}{Nature} \textbf{\bibinfo{volume}{396}},
  \bibinfo{pages}{345} (\bibinfo{year}{1994}).

\bibitem[{\citenamefont{Ho}(1998)}]{ho1998}
\bibinfo{author}{\bibfnamefont{T.-L.} \bibnamefont{Ho}},
  \bibinfo{journal}{Phys. Rev. Lett.} \textbf{\bibinfo{volume}{81}},
  \bibinfo{pages}{742} (\bibinfo{year}{1998}).

\bibitem[{\citenamefont{Ohmi and Machida}(1998)}]{ohmi1998}
\bibinfo{author}{\bibfnamefont{T.}~\bibnamefont{Ohmi}} \bibnamefont{and}
  \bibinfo{author}{\bibfnamefont{K.}~\bibnamefont{Machida}},
  \bibinfo{journal}{J. Phys. Soc. Jpn.} \textbf{\bibinfo{volume}{67}},
  \bibinfo{pages}{1822} (\bibinfo{year}{1998}).

\bibitem[{\citenamefont{Murata et~al.}(2007)\citenamefont{Murata, Saito, and
  Ueda}}]{murata2006}
\bibinfo{author}{\bibfnamefont{K.}~\bibnamefont{Murata}},
  \bibinfo{author}{\bibfnamefont{H.}~\bibnamefont{Saito}}, \bibnamefont{and}
  \bibinfo{author}{\bibfnamefont{M.}~\bibnamefont{Ueda}},
  \bibinfo{journal}{Phys. Rev. A} \textbf{\bibinfo{volume}{75}},
  \bibinfo{pages}{013607} (\bibinfo{year}{2007}).

\bibitem[{\citenamefont{Mukerjee et~al.}(2006)\citenamefont{Mukerjee, Xu, and
  Moore}}]{mukerjee2006}
\bibinfo{author}{\bibfnamefont{S.}~\bibnamefont{Mukerjee}},
  \bibinfo{author}{\bibfnamefont{C.}~\bibnamefont{Xu}}, \bibnamefont{and}
  \bibinfo{author}{\bibfnamefont{J.~E.} \bibnamefont{Moore}},
  \bibinfo{journal}{Phys. Rev. Lett} \textbf{\bibinfo{volume}{97}},
  \bibinfo{eid}{120406} (\bibinfo{year}{2006}).

\bibitem[{\citenamefont{Lamacraft}(2004)}]{lamacraft2006}
\bibinfo{author}{\bibfnamefont{A.}~\bibnamefont{Lamacraft}},
  \bibinfo{journal}{Unpublished}  (\bibinfo{year}{2004}).

\bibitem[{\citenamefont{Fisher et~al.}(1989)\citenamefont{Fisher, Weichman,
  Grinstein, and Fisher}}]{fisher1989}
\bibinfo{author}{\bibfnamefont{M.~P.~A.} \bibnamefont{Fisher \emph{et al.}}},
 \bibinfo{journal}{Phys. Rev. B}
  \textbf{\bibinfo{volume}{40}}, \bibinfo{pages}{546} (\bibinfo{year}{1989}).

\bibitem[{\citenamefont{Weinberg and Wu}(1987)}]{weinberg1987}
\bibinfo{author}{\bibfnamefont{E.~J.} \bibnamefont{Weinberg}} \bibnamefont{and}
  \bibinfo{author}{\bibfnamefont{A.}~\bibnamefont{Wu}}, \bibinfo{journal}{Phys.
  Rev. D} \textbf{\bibinfo{volume}{36}}, \bibinfo{pages}{2474}
  (\bibinfo{year}{1987}).

\bibitem[{\citenamefont{Guth and Pi}(1985)}]{guth1985}
\bibinfo{author}{\bibfnamefont{A.~H.} \bibnamefont{Guth}} \bibnamefont{and}
  \bibinfo{author}{\bibfnamefont{S.-Y.} \bibnamefont{Pi}},
  \bibinfo{journal}{Phys. Rev. D} \textbf{\bibinfo{volume}{32}},
  \bibinfo{pages}{1899} (\bibinfo{year}{1985}).

\bibitem[{\citenamefont{Halperin}(1981)}]{halperin1981}
\bibinfo{author}{\bibfnamefont{B.~I.} \bibnamefont{Halperin}}, in
  \emph{\bibinfo{booktitle}{Physics of Defects}}, edited by
  \bibinfo{editor}{\bibfnamefont{R.}~\bibnamefont{Balian}}, \bibinfo{editor}{\bibfnamefont{M.}~\bibnamefont{Kleman}} \bibnamefont{and}
  \bibinfo{editor}{\bibfnamefont{J.~P.}~\bibnamefont{Poirier}} 
  (\bibinfo{publisher}{North-Holland
  Press}, \bibinfo{year}{1981}).

\bibitem[{\citenamefont{Liu and Mazenko}(1992)}]{liu1992}
\bibinfo{author}{\bibfnamefont{F.}~\bibnamefont{Liu}} \bibnamefont{and}
  \bibinfo{author}{\bibfnamefont{G.~F.} \bibnamefont{Mazenko}},
  \bibinfo{journal}{Phys. Rev. B} \textbf{\bibinfo{volume}{46}},
  \bibinfo{pages}{5963} (\bibinfo{year}{1992}).

\bibitem[{\citenamefont{Hohenberg and Halperin}(1977)}]{hohenberg1977}
\bibinfo{author}{\bibfnamefont{P.~C.} \bibnamefont{Hohenberg}}
  \bibnamefont{and} \bibinfo{author}{\bibfnamefont{B.~I.}
  \bibnamefont{Halperin}}, \bibinfo{journal}{Rev. Mod. Phys.}
  \textbf{\bibinfo{volume}{49}}, \bibinfo{pages}{435} (\bibinfo{year}{1977}).

\bibitem[{\citenamefont{Kockelkoren and Chat\'e}(2002)}]{kockelkoren2002}
\bibinfo{author}{\bibfnamefont{J.}~\bibnamefont{Kockelkoren}} \bibnamefont{and}
  \bibinfo{author}{\bibfnamefont{H.}~\bibnamefont{Chat\'e}},
  \bibinfo{journal}{Phys. Rev. E} \textbf{\bibinfo{volume}{65}},
  \bibinfo{pages}{058101} (\bibinfo{year}{2002}).

\bibitem[{\citenamefont{Damle et~al.}(1996)\citenamefont{Damle, Majumdar, and
  Sachdev}}]{damle1996}
\bibinfo{author}{\bibfnamefont{K.}~\bibnamefont{Damle}},
  \bibinfo{author}{\bibfnamefont{S.~N.} \bibnamefont{Majumdar}},
  \bibnamefont{and} \bibinfo{author}{\bibfnamefont{S.}~\bibnamefont{Sachdev}},
  \bibinfo{journal}{Phys. Rev. A} \textbf{\bibinfo{volume}{54}},
  \bibinfo{pages}{5037} (\bibinfo{year}{1996}).

\bibitem[{\citenamefont{Saito et~al.}(2006)\citenamefont{Saito, Kawaguchi, and
  Ueda}}]{saito2006b}
\bibinfo{author}{\bibfnamefont{H.}~\bibnamefont{Saito}},
  \bibinfo{author}{\bibfnamefont{Y.}~\bibnamefont{Kawaguchi}},
  \bibnamefont{and} \bibinfo{author}{\bibfnamefont{M.}~\bibnamefont{Ueda}}
    \bibinfo{journal}{Phys. Rev. A} \textbf{\bibinfo{volume}{75}},
  \bibinfo{pages}{013621} (\bibinfo{year}{2007}).

\end{thebibliography}
 \end{document}